\documentstyle[bo99,epsfig]{article}
\def\ltsim{\ifmmode\stackrel{<}{_{\sim}}\else$\stackrel{<}{_{\sim}}$\fi}
\def\gtsim{\ifmmode\stackrel{>}{_{\sim}}\else$\stackrel{>}{_{\sim}}$\fi}

\title{Observing the Unobservable: Coronal \\ Line Data from Narrow-Line
Seyfert~1s as \\ a Test of EUV Continuum Models }
\author{P.J. Bleackley and P.T.O'Brien}
\affil{Department of Physics and Astronomy, University of Leicester }

\begin{document}

\maketitle

\begin{abstract}
We have measured coronal line emission for a sample of 19 Narrow-Line
Seyfert~1 galaxies. Kinematic analysis shows that the Coronal Line
Region (CLR) consists of gas in a decelerating outflow.

We intend to use these data as a test of the EUV emission of NLS1s. By
comparison with photoionization modelling for a number of
power-law + black body spectra, we plan to deduce the most likely EUV
continua for these objects as well as explore physical conditions in
the CLR.
\end{abstract}

\section{Introduction }
Narrow-line Seyfert~1 objects (NLS1s) are a phenomenologically interesting
subclass of Seyfert~1s. As their name suggests, they have the narrowest
permitted lines of all Seyfert~1s (FWHM(H${\beta}$) $\ltsim$ 2000 km
s$^{-1}$), which appear to be associated with a range of other extremal
properties, including strong, steep, soft X-ray excesses, strong permitted
Fe~II lines, weak forbidden lines, and a high degree of short timescale
variability in X-rays. Coronal lines, forbidden transitions from highly
ionised states of iron, are often present in their optical spectra. Their
properties form a continuum with those of ``normal'' Seyfert 1s, but as
they show extremal properties, they are important for testing AGN
unification schemes. It is thought likely that they are Seyfert 1s with
low mass black holes and/or high accretion rates.

To further understand the nature of the soft UV/X-ray excess, which is an
important component of the energy output of these objects, we wish to
study the EUV continuum. As this is not directly observable for
extragalactic objects due to Galactic absorption, we use coronal lines
(which have ionisation energies $\sim$ a few hundred eV) as diagnostics of
this region. This approach also allows us to investigate the physical
conditions in the coronal line region, particularly by kinematic methods. 

\section{Observations and data reduction }

Nineteen NLS1s from the soft X-ray observed sample of Boller et al.  
(1996)  were observed between 5 -- 11 August 1996, using the Intermediate
Dispersion Spectrograph mounted on the Isaac Newton Telescope at La Palma,
obtaining dispersions of 1.6 -- 3.2 \AA \ pixel$^{-1}$. The data were
reduced with the Starlink project's \texttt{FIGARO} software, and fitted
using the \texttt{ELF} routines in \texttt{DIPSO}, which simultaneusly
fitted gaussian line profiles and a linear continuum. This approach
allowed the deblending of neighbouring lines.

The coronal lines [Fe VII] $\lambda$6038.9, [Fe X] $\lambda$6374.6, [Fe
XI] $\lambda$7891.9 and [Fe XIV] $\lambda$5302.86 were identified where
possible, and upper limits were placed on them otherwise. Line profiles
were then transformed into velocity space relative to the rest-frame of
the [O III] $\lambda\lambda$4959, 5007 doublet.

\section{Results}

\begin{figure}[h!]
\centerline{\psfig{file=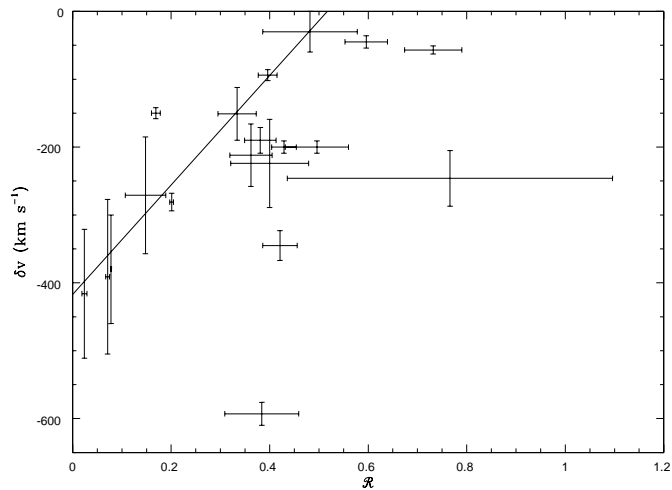, width=7cm, angle=270}}
\caption[]{The decelerating outflow of the Coronal Line Region}
\end{figure}

\subsection{Kinematic analysis }
Erkens et al. (1997) noted a tendency for broader coronal lines to be more
blueshifted than narrower ones. To further investigate this
correlation in a manner independent of black hole mass, we assume that
the FWHM of coronal lines is governed by the virial theorem and define a 
\textit{radial estimator}
\[ \mathcal{R} = \frac{\mathit{r}_{\mbox{\tiny
[Fe]}}}{\mathit{r}_{\mbox{\tiny [O III]}}} = \left(
\frac{\mbox{FWHM}_{\mbox{\tiny [O III]}}}{\mbox{FWHM}_{\mbox{\tiny [Fe]}}}
\right)^{\mathrm{2}} \]
where $r_{\mbox{\tiny [Fe]}}$ is the distance from the central black hole
at which a line, [Fe], is emitted. Plotting the outflow velocity $\delta
v$ against $\mathcal{R}$ shows a decelerating outflow, which was fitted to
first order by $\delta v = (806 \pm 93)\mathcal{R} - \rm (417 \pm 25)$ km
s$^{-1}$ (see Figure 1). The minimum value of $\mathcal{R}$ observed is
0.024, which constrains the maximum size of the region over which the gas
is accelerated.

For those objects for which at least 3 coronal lines had been detected, we
plotted $\mathcal{R}$ against the ionisation potential, in order to
determine if stratification of the CLR was important. For Arakelian 564,
there is a trend towards more highly ionised gas being detected closer to
the nucleus, but this trend is not unambiguously detected in other objects
(see Figure 2). 

\begin{figure}[h!]
\centerline{\psfig{file=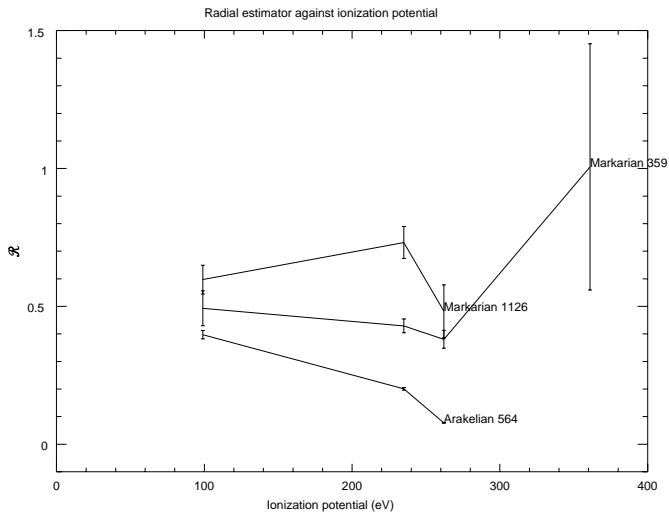, width=7cm, angle=270}}
\caption[]{Radial estimator against ionisation potential for coronal
lines.}
\end{figure}

\subsection{Comparison with X-ray data}
\begin{figure}[h!]
\centerline{\psfig{file=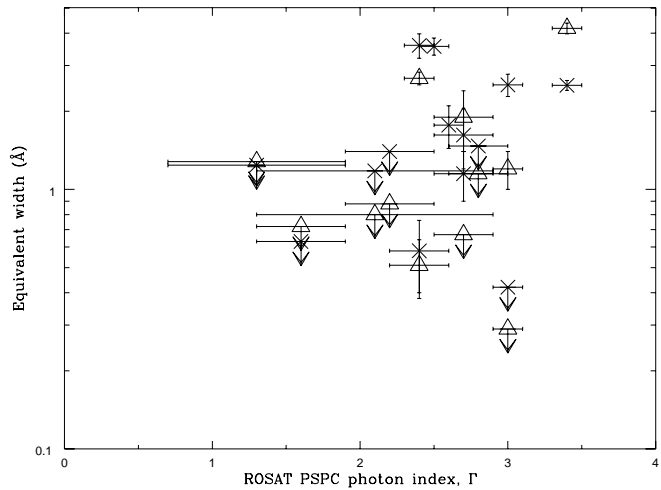, width=7cm, angle=270}}
\caption[]{Equivalent width of [Fe VII] (crosses) and [Fe X] (triangles) 
against ROSAT PSPC photon index.}
\end{figure}

Equivalent widths of [Fe VII] and [Fe X] were compared with ROSAT PSPC
photon indices obtained from Boller et al. (1996) (see Figure 3). No
strong correlation is present for either line. 

\section{Conclusions}

Gas in the coronal line region is outflowing, and this outflow appears to
be decelerating. The coronal line gas has reached an outflow velocity of
420 km s$^{-1}$ within a distance from the nucleus estimated to be $<1/40$
that of the NLR. Radiation pressure does not appear to drive motion in the
CLR, as this would be expected to produce accelerating, rather than
decelerating outflows (see Binette (1998) for an example of a radiation
pressure driven model).

Coronal line strengths do not follow a simple correlation with ROSAT PSPC
photon index. To investigate the EUV continuum in more detail, we plan
to create photoionisation models based on a fixed power-law, with a black
body component of variable luminosity and temperature added, and test  the
predictions of these models against our observed coronal line strengths.
The models will also allow for stratification of the CLR.

\begin{acknowledgements}
We thank Dr. T. Roberts for his help with the observations. This work uses
data obtained using the Isaac Newton Telescope, Observatorio de la Roque
de los Muchachos, La Palma. P.J. Bleackley has been supported by a PPARC
research studentship. 
\end{acknowledgements}


\begin{references}

\reff Binette, L., 1998, MNRAS, 294, L47--L51

\reff Boller, Th., Brandt, W.N., and Fink, H., 1996, A\&A, 305, 53--73

\reff Erkens, U., Appenzeller, I., and Wagner, S., 1997, A\&A, 323,
707--716 

\end{references}
\end{document}